\documentclass[aps,pre,preprint,groupedaddress,showpacs]{revtex4-1}

\usepackage{amssymb}
\usepackage{amsmath}
\usepackage{floatrow}
\usepackage{subfigure}
\usepackage{subfloat}
\usepackage{graphicx}

\usepackage[colorlinks,citecolor=blue]{hyperref}

\begin{document}


\title{Breather and Rogue Wave  solutions of a Generalized Nonlinear Schr\"odinger Equation}


\author{ L.H. Wang $^{1}$, K. Porsezian $^{2}$ and J.S. He $^{1}$  }

\email[Corresponding author: ]{hejingsong@nbu.edu.cn}

\affiliation{
$^{1}$Department of Mathematics, Ningbo University,  Ningbo , Zhejiang 315211, P.\ R.\ China \\
$^{2}$Department of Physics, Pondicherry University, Puducherry 605014, India
}


\date{\today}

\begin{abstract}
In this paper, using the Darboux transformation, we
demonstrate the generation of first-order breather and higher-order
rogue waves from a generalized nonlinear Schr\"odinger equation with
several higher-order nonlinear
effects representing femtosecond pulse propagation through nonlinear
 silica fiber. The same nonlinear evolution equation can also describes the soliton-type nonlinear excitations in classical Heisenberg spin chain. Such
solutions have a parameter $\gamma_1$, denoting the strength of the
higher-order effects. From the numerical plots of the rational
solutions, the compression effects of the breather and rogue waves
produced by $\gamma_1$ are discussed in detail.
\end{abstract}


\maketitle

\section{Introduction}
\label{}

It is well known that one of the most challenging aspects of modern
science and technology is the nonlinear nature of the system, which
is considered to be fundamental to the understanding of many natural
phenomena. In recent years, nonlinear science has emerged as a
powerful subject for explaining the mysteries of the challenging
nature. Nonlinearity is a fascinating occurrence of nature whose
importance has been well appreciated for many years, in the context
of large amplitude waves or high-intensity laser pulses observed in
various fields ranging from fluids to solid state, chemical,
biological,  nonlinear optical, and geological systems \cite{GPA1,
AH1, LFM, KP1, KP2, JRT1, FA1, FA2, YSK}. This fascinating subject
has branched out in almost all areas of science, and its
applications are percolating through the whole of science. In
general, nonlinear phenomena are often modeled by nonlinear
evolution equations exhibiting a wide range of high complexities in
terms of different linear and nonlinear effects. In recent years,
the advent of high-speed computers, many advanced mathematical
software, and development of  many sophisticated and systematic
analytical methods in the study of the nonlinear phenomena and also
supported by many experiments have encouraged both theoretical and
experimental research. In the past few decades, nonlinear
science has experienced an explosive growth by the invention of
several exciting and fascinating new concepts, such as solitons,
dispersion-managed solitons, dromions, rogue waves, similaritons,
supercontinuum generation, complete integrability, fractals, chaos,
etc. \cite{GPA1, AH1, LFM, KP1, KP2, JRT1, FA1, FA2, YSK, GPA2,
NN1}. Many of the completely integrable nonlinear partial
differential equations (NPDEs) admit one of the most striking
aspects of nonlinear phenomena called soliton, which describe
soliton as a universal character, and they are of great mathematical
as well as physical interest, too. The study of the solitons and
other related issues of the construction of the solutions to a wide
class of NPDEs have become one of the most exciting and extremely
active areas of research in science and technology for many years.

In addition to several developments in soliton theory, recent
developments in modulational instability (MI) have also been
widely used to explain why experiments involving white coherent
light supercontinuum generation (SCG), admit a triangular spectrum.
Such universal triangular spectra can be well described by the
analytical expressions for the spectra of Akhmediev breather
solutions at the point of extreme compression. In the context
of the NLS equation, Peregrine already in Ref. \cite{peregrine} had
identified the role of MI in the formation of patterns
resembling freak waves or rogue wave (RW); these theoretical
results were later supported by several experiments. Rogue
waves in the ocean are localized large amplitude waves on a
rough background, which have two remarkable
characteristics: (1) ``appear from nowhere and
disappear without a trace" \cite{akhmediev1},
(2) exhibit one dominant peak. RWs have recently
appeared in several areas of science. Particularly
in photonic crystal fibers, RWs have been well
established in connection with SCG \cite{soli1}.
This actually has stimulated research for RWs
in other physical systems and has paved the way
for many important applications, including the control of RWs by
means of SCG \cite{soli2,dudley}, as well as studies in superfluid
Helium \cite{ganshin}, Bose-Einstein condensates \cite{konotop1},
plasmas \cite{ruderman, shukla}, microwave \cite{hohmann}, capillary
phenomena \cite{shats}, in telecommunication data
streams \cite{vergeles}, inhomogeneous
media \cite{arecchi}, water
experiments \cite{akhmediev3, akhmediev4}, and so on.
More recently, Kibler \emph{et al.} \cite{akhmediev2},
using their elegant experimental apparatus in optical fibers, were able
to generate femtosecond pulses with strong temporal and spatial
localization and near-ideal temporal Peregrine soliton characteristics.

In the recent past, several equations have been shown to admit the
rogue wave solutions. For example, in addition to the NLS equation,
the Hirota equation \cite{akhmediev5,he1,lu1}, the first-type
derivative NLS equation \cite{he2}, the third-type DNLS equation
\cite{he3}, the Fokas-Leneels equation \cite{xuhepfoakseq}, the
NLS-MB equations \cite{he4}, the Hirota Maxwell-Bloch(MB) equations
\cite{lihephmb}, the Sasa-Satsuma equation \cite{Bandelow},  the
discrete Ablowitz-Ladik and Hirota equation \cite{akhmediev6}, the
two-component NLS equations \cite{akhmediev6b,guo1,fabio1}, the
three-components NLS equations \cite{qin1}, the variable coefficient
NLS \cite{akhmediev7, taki1,yan1,he5,liu1}, the variable coefficient
derivative NLS \cite{he6}, and the variable coefficient higher-order
NLS (VCHNLS) \cite{zhang1}, the rogue waves in dissipative systems
\cite{Philippe} are a few of the nonlinear evolution equations that
admit RWs. From the above studies, it is clear that one of the
possible generating mechanisms \cite{hezhangwangpf2012} for the
higher-order RW is the interaction of the multiple breathers
possessing the same and the very particular frequency of the
underlying equation.

In recent years, there has been a considerable interest in the study of the nonlinear excitations of the spin chains with competing bilinear
and biquadratic interactions. In particular, the complete
integrability and nonlinear excitations of spin chains with
spin magnitude $S>1$ has been established if suitable
polynomials in ($S_i,S_j$) are added to the original bilinear
Heisenberg spin Hamiltonian. In this connection and also
from the mathematical point of view, it is of interest
to study the influence of the biquadratic interactions
on the integrability of the Heisenberg bilinear spin
chain in the classical limit as well. Considering the
above points, one of the authors of this paper, has
investigated the integrability aspects of a classical
one-dimensional isotropic biquadratic Heisenberg
spin chain in its continuum limit up to order $O(a^4)$
in the lattice parameter through a classical differential
geometric approach \cite{kpspin1,
kpspin2, kpspin3} and investigated the soliton and
integrability aspects of the corresponding generalized nonlinear
Schr\"odinger Equation (GNLSE). This equation is given by \cite{kpspin1}
\begin{equation}\label{e1}
i q_{{t}}  +  q_{{xx}}+2{q} |q|^2  +
 \gamma_1 \left( q_{{xxxx}}+6{q_x^2}q^*+4q |q_x|^2+8 |q|^2 q_{xx}
+2{q}^{2}q^*_{{xx}}+6|q|^4 {q}\right) =0.
\end{equation}
Here, $q(x,t)$ is the complex envelope and $\gamma_1$ denotes the
strength of higher-order linear and nonlinear effects.
 When we consider the propagation of ultra short pulse propagation through optical fiber, i.e, less than 100-fs pulses, it has been shown that higher-order dispersion, self-steepening, self-frequency, and quintic effects should be included in the model.  The above equation has been shown to be integrable and admits exact soliton solutions and also gauge equivalent to Heisenberg spin chain equation. Thus, it is an interesting problem to find how these higher-order
terms will affect the breather and  rogue wave in an associated optical
system and spin system by means of  changing the value of $\gamma_1$. It is
our prime aim to answer this problem in this paper.

The paper is organized as follows.  In Sec. II, the Lax pair and
the Darboux transformation (DT) are introduced. In Sec. III, we shall
give the first-order breather and it's limit of infinitely large
period. In Sec. IV, higher-order rogue waves are given.
The compression effects on the breather and rogue waves
produced by the higher-order terms of GNLSE are given in
Secs. III and IV, respectively. Section V is devoted
to conclusions.

\section{ Lax Pair and Darboux transformation}

As discussed above, in this section, we would like to recall the Lax
pair of GNLSE \cite{kpspin1, kpspin2, kpspin3}and to show its
Darboux transformation. According  to the AKNS formalism,  Lax pair
for Eq. (\ref{e1}) is written as
\begin{equation}\label{e2}
\Phi_x=M\Phi, \Phi_t=N\Phi
\end{equation}
with the following matrices:
$$
M=i\lambda U_0+U_1=
i\lambda
\left(
\begin {array}{cc}
-1&0 \\
0  & 1
\end {array}
\right)
+
\left(
\begin {array}{cc}
0&q \\
-q^*  &0
\end {array}
\right)
=
\left(
\begin {array}{cc}
-i\lambda&q \\
-q^*  & i\lambda
\end {array}
\right),
$$
and $N=8i\gamma_1V_4-2i V_2$. Here,
\[
V_2=
\left(
\begin {array}{cc}
{\lambda}^{2}-\frac{1}{2}\,qq^* &  iq\lambda-\frac{1}{2}q_{{x}}\\
-iq^*\,\lambda-\frac{1}{2}\,q^*_{{x}}&-{\lambda}^{2}+\frac{1}{2}\,qq^*
\end {array}
\right),  V_4= \left(
\begin {array}{cc}
A_4&B_4 \\
C_4&-A_4
\end {array}
\right),
\]
\[
\begin{split}
A_4= &  {\lambda}^{4}-\frac{1}{2} qq^* {\lambda}^{2}+ \frac{i}{4}\left(
qq^*_{{x}}-q_{{x}} q^*\right) \lambda+\frac{1}{8}(3
{q}^{2}{q^*}^{2}+ q^*\,q_{{x,x}}+ qq^*_{{x,x}}- q_{{x}}q^*_{{x}}),
\\
B_4= &  iq{\lambda}^{3}-\frac{1}{2}q_{{x}}{\lambda}^{2}- \frac{i}{4}\left( q_{{x,x}}+2{q}^{2}q^*
 \right) \lambda+ \frac{1}{8}(q_{{x,x,x}}+6\,qq^*\,q_{{x}}),
\\
C_4= &  -iq^*\,{\lambda}^{3}-\frac{1}{2}q^*_{{x}}{\lambda}^{2}+
\frac{i}{4}\left(  q^*_{{x,x}} +2 q{q^*}^{2} \right)
\lambda+\frac{1}{8} (q^*_{{x,x,x}}+6qq^*\,q^*_{{x}} ).
\end{split}
\]
Moreover,
$$
\Phi(\lambda)= \left(
\begin {array}{c}
\phi(\lambda)
\\
\psi(\lambda)
\end {array}
\right)= \left(
\begin {array}{c}
\phi(\lambda;x,t)
\\
\psi(\lambda;x,t)
\end {array}
\right)$$
 denotes the eigenfunction of Lax pair Eq. (\ref{e2})
associated with $\lambda$.

The Lax pair of GNLSE provides a basis for the solvability of
this equation by means of the Darboux transformation. To
construct the $n$-fold DT, it is necessary to introduce $2n$
eigenfunctions $f_i= \left(
\begin {array}{c}
f_{i,1}
\\
f_{i,2}
\end {array}
\right)=\Phi(\lambda_i)$, associated with eigenfunction
$\lambda_i(i=1,2,\cdots, 2n)$, and satisfy corresponding reduction
condition $\lambda_{2k}=\lambda^*_{2k-1}$  as we have done for the
NLS equation \cite{Matveev,Hedeterminant}. Furthermore, a similar
$n$-fold DT determinant representation derived for the NLS equation as
given in \cite{Matveev}. 
 For example, from the one-fold DT, we get
\begin{equation}\label{e3}
q^{[1]}=q^{[0]}-{\frac {2\,i\Delta_{{1}} }{\Delta_{{2}}}},
\end{equation}
with
$\lambda_{{1}}=\xi+i\eta,\lambda_{{2}}=\lambda_{{1}}^*=\xi-i\eta,f_{{2,1}}=-f_{{1,2}}^*,
f_{{2,2}}=f_{{1,1}}^*, \Delta_{{1}}= \left|
\begin {array}{cc}
f_{{1,1}}&\lambda_{{1}}f_{{1,1} }
\\
f_{{2,1}}& \lambda_{{2}}f_{{2,1}}
\end {array}
\right| = -2\,if_{{1,1}}f_{{2,1}}\eta,
 \Delta_{{2}}= \left|
\begin {array}{cc}
f_{{1,1}}&f_{{1,2}}
\\
f_{{2,1}}&f_{{2,2}}
\end {array}
\right| = f_{{1,1}}f_{{2,2}}-f_{{1,2}}f_{{2,1}}.
$

\section{The first-order breather and its limit}
In this section, we first solve the eigenfunctions associated with a
periodic seed $q^{[0]}$,  and then use it to get a first-order
breather by using the  determinant representation of one-fold DT in
Eq. (\ref{e3}). Further, this breather implies a first-order rogue
wave in the limit of infinitely large period. These two solutions have
$\gamma_1$ explicitly such that we can use it to study the
effects of breather and rogue waves affected by the higher-order terms.

Considering a periodic solution in the following form:
\begin{equation}\label{e6}
 q^{[0]}=c{{\rm e}^{i  \rho  }}
\end{equation}
with $\rho=ax+bt, b= \left( {a}^{4} -12{a}^{2}{c}^{2}+6{c}^{4}
\right) \gamma_{{1}}+ 2\,{c}^{2}-{a}^{2}$.  By the  method of
separation of variables and the superposition principle, we have the
following eigenfunction associated with $q^{[0]}$:
\begin{eqnarray}\label{eqf11}
&f_{1,1}=&k_{{1}}c{{\rm e}^{i \left( \frac{\rho}{2}+d \right)}}
+ik_{{2}} \left( \frac{a}{2}+h+\lambda \right) {{\rm e}^{i \left(
\frac{\rho}{2}-d \right) }},\\
&f_{1,2}=&k_{{2}}c{{\rm e}^{-i \left( \frac{\rho}{2}+d \right) }}
+ik_{{1}} \left( \frac{a}{2}+h+\lambda \right) {{\rm e}^{-i \left(
\frac{\rho}{2}-d \right) }}, \label{eqf12}
\end{eqnarray}
with, $h=\sqrt{{c}^{2}+ \left( \lambda+\frac{a}{2} \right) ^{2}}=h_R + i h_I$,$k_1={{\rm e}^{ih \left( s_1\,\epsilon+s_2\,{\epsilon}^{ 2} \right) }}$,
$k_2={{\rm e}^{-ih \left( s_1\,\epsilon+s_2\,{\epsilon}^{ 2} \right) }}$,
\begin{eqnarray*}
&d&=\gamma_{1} \left( a \left( {a}^{2}-6\,{c}^{2}
\right)-8\,{\lambda}^{3}+4\,a{\lambda}^{2}+ \left(
4\,{c} ^{2}-2\,{a}^{2} \right) \lambda \right) ht + \left(x+\left(2\lambda - a \right)t
\right)h \\
& &= \left(x+\left(2\lambda - a + \gamma_{1} \left( a \left( {a}^{2}-6\,{c}^{2}
\right)-8\,{\lambda}^{3}+4\,a{\lambda}^{2}+ \left(
4\,{c} ^{2}-2\,{a}^{2} \right) \lambda \right)  \right)t
\right)h \\
& &=(x+( d_R+i d_I)t) h
\end{eqnarray*}

Using the one-fold DT, a first-order breather is constructed in the form
    {\footnotesize
\begin{equation}\label{1breather}
q^{[1]}=\mbox{\hspace{-0.2cm}}\left(c+\frac
{2\eta\left\{\left[w_1\cos\left(2G\right)-w_2\cosh\left(2F\right)
\right]-i\left[\left(w_1-2c^2\right)\sin\left(2G\right)- w_3\sinh
\left(2F\right)\right]\right\}}{w_1\cosh\left(2F\right)-w_2\cos
\left(2G\right)}\right){\rm e}^{i\rho},
\end{equation}
} with $w_1={c}^{2}+(h_I+\eta)^{2}+(
\xi+h_R+\frac{a}{2})^{2},w_2=2c(h_I+\eta),
w_3=2c(\xi+h_R+\frac{a}{2})$, $F=  x h_I +(d_R h_I + d_I h_R) \,t,G=
x h_R +(d_R   h_R - d_I h_I)t$.  This is a periodic traveling wave.
The coefficient $\gamma_1$ can affect the period of the breather
through $G$.

It is trivial to find
$$|q^{[1]}|^2(0,0)=(c+2\eta)^2,$$ which is the height of peaks of this breather. Obviously,
the height is independent of  $a,\xi$ and $\gamma_1$. This does not mean
that $\gamma_1$ cannot affect the properties of the breather. In
fact, we can see from Eq. (\ref{1breather}) that $\gamma_1$ actually controls
the period of the breather. This observation can be
clearly seen in Fig. \ref{fig2}: the number of peaks on same
time interval is increasing when $\gamma_1$ goes up from $0$ to $3$
with a constant gap $0.5$ (we skip the figure for $\gamma_1=\frac{5}{2}$ ) . These pictures clearly show that
the resulting  breather is compressed by the higher-order effects due to the presence of  $\gamma_1$ than $\gamma_1=0$ case. In addition to the above, when the value of $\gamma_1$ increases, the number of peaks also increases. We use a short interval in Fig. \ref{fig2}(a) and \ref{fig2}(b) to avoid too many peaks
in it.

  Now we can consider what will happen in a breather when its period
goes to infinity. According to the explicit expression in
Eq. (\ref{1breather}) of the first-order breather,
this limit can be realized  by setting
$\displaystyle \lambda= \xi+i\eta \rightarrow \lambda_0
=-\frac{a}{2}  + i c $; i.e., $\lim_{\lambda \rightarrow
\lambda_0 }{q^{[1]}}$. For simplicity, on setting $a=0$ ,
the limit of the breather solution is obtained as
\begin{equation}\label{e9}
q^{[1]}_{limit}=
\left( \frac{4(1+iT)}{ X^2+T^2 + 1} - 1 \right)
c {{\rm e}^{i 2 {c}^{2} \left( 3\,\gamma_{{1}}{c}^{2}+1 \right) t}},
\end{equation}
with $T=4\left( 1+6\,\gamma_{{1}}{c}^{2} \right) {c}^{2} t , \quad
X=2 c x $. This is nothing but a first-order rogue wave possessing
asymptotic height $1$ when $x$ and $t$ go to infinity. Further, we
find $|q^{[1]}_{limit}|_{max}^2(0,0)=9 c^2$, which denotes the
height of a first-order RW.  We can see from $T$ that, like in the case of breather compression,  $\gamma_1$ is
also responsible for compression effect of RW in the time direction,
which is clearly seen in Fig. \ref{fig4} with
$\gamma_1=0,0.5,1,2$, respectively.
\section{Higher-order Rogue Waves}
The limit method in Eq. (\ref{e9}) is not applicable for the
higher-order breather when
$\lambda_i \rightarrow \lambda_0(i \geq 2)$.
We can overcome this problem by using the coefficient of the
Taylor expansion in the determinant representation of a
higher-order breather $q^{[n]}$ \cite{ hezhangwangpf2012,Hedeterminant}. Similar to the case of NLS \cite{hezhangwangpf2012}, the first-order rogue wave of GNLSE is given by

\begin{equation}\label{e10}
q^{[1]}_{rw}=
\left( \frac{F_1+i G_1}{H_1} - 1 \right) c{{\rm e}^{i\rho}}.
\end{equation}
Here,
$F_1=4, \quad
G_1=16 \left( 1+6\,\gamma_{{1}}{c}^{2}-6\,{a}^{2}\gamma_{{1}}
\right) {c}^{2} t,
$
\begin{equation*}
\begin{split}
 H_1 =& 4\,{c}^{2}{x}^{2} + \left( 32\,{c}^{2}\gamma_{{1}}{a}^{3}-16\,{c}^{2} \left( 1+12\,\gamma_{{1}}{c}^{2} \right) a \right) x t
 +( 64\,{c}^{2}{\gamma_{{1}}}^{2}{a}^{6} \\
      & -64\,{c}^{2}\gamma_{{1}} \left( 1+3\, \gamma_{{1}}{c}^{2} \right) {a}^{4} +16\,{c}^{2} \left( 1+12\,\gamma_{{
1}}{c}^{2}+72\,{c}^{4}{\gamma_{{1}}}^{2} \right) {a}^{2}\\
      & +16\,{c}^{4}
\left( 6\,\gamma_{{1}}{c}^{2}+1 \right) ^{2} ) {t}^{2} + 1.
\end{split}
\end{equation*}
Note that $q^{[1]}_{rw}$ reduces to the $q^{[1]}_{limit}$
when $a=0$.

What follows is the second-order rogue wave given by the Taylor
expansion when $\lambda_i \rightarrow \lambda_0(i=1,3)$
as the case of NLS \cite{hezhangwangpf2012}. There are two patterns
for the second-order RW. The first one is called the fundamental pattern
possessing a highest peak surrounded by four small equal peaks in
two sides. Setting $a=0, s_1=0$, then the Taylor expansion in
the determinant of $q^{[2]}$ provides
\begin{equation}\label{e11}
q^{[2]}_{rw}= \left( \frac{12(F_2 + i G_2)}{H_2} - 1 \right) c{{\rm
e}^{i 2 {c}^{2} \left( 3\,\gamma_{{1}}{c}^{2}+1 \right) t}}.
\end{equation}
Here,
\begin{equation*}
\begin{split}
F_2=&5\,{T}^{4}+ \left( 6\,{X}^{2}+34 \right) {T}^{2}-64\,Tt{c}^{2}+
{X}^{4}+6\,{X}^{2}-3,  \\
G_2=&{T}^{5}+ \left( 2\,{X}^{2}+10 \right) {T}^{3}-32\,t{c}^{2}{T}^{2}+
 \left({X}^{4}-14\,{X}^{2}-23 \right) T+32\,t{c}^{2} \left( {X}^{2}+1 \right),\\
H_2=& {T}^{6}+ \left( 3\,{X}^{2}+43 \right) {T}^{4}-64\,t{c}^{2}{T}^{3}+
       \left( 3\,{X}^{4}-66\,{X}^{2} +307 \right) {T}^{2} \\
    & + \left( 192\,{c}^{2}t{X}^{2}-1088\,t{c}^{2} \right) T+1024\,{c}^{4}{t}^{2}+{X}^{6}+3\,{X}^{4}+ 27\,{X}^{2}
     + 9,
\end{split}
\end{equation*}
and $X,T$ are defined in Eq. (\ref{e9}). By comparing two cases with
$\gamma_1=0$ and $\gamma_1=1$ in Figure \ref{fig5}, the compression
effect in $t$ direction is shown clearly. The second is a triangular
pattern, which consists of three equal peaks. Setting
$s_2=0$,$a=0,s_1=50-50i$, then an explicit formula of this pattern
is
\begin{equation}
q^{[2]}_{rwtrig}=\left( 1- \frac{12(F_{2trig} + i G_{2trig})}{H_{2trig}} \right) c{{\rm e}^{i 2 {c}^{2} \left( 3\,\gamma_{{1}}{c}^{2}+1 \right) t}} ,
\end{equation}
with
\begin{eqnarray*}
&F_{2trig}=& 5\,{T}^{4}+ \left( 24\, \left( 1+6\,\gamma_{{1}}{c}^{2} \right) t{c}^{
2}{X}^{2}+24\, \left( 34\,\gamma_{{1}}{c}^{2}+3 \right) t{c}^{2}+1200
\,{c}^{2} \right) T  \\
& & -3+1200\,{c}^{2}X+6\,{X}^{2}+{X}^{4} ,\\
&G_{2trig}=&  {T}^{5}+ \left( 8\, \left( 1+6\,\gamma_{{1}}{c}^{2} \right) t{c}^{2}{X
}^{2}+8\, \left( 1+30\,\gamma_{{1}}{c}^{2} \right) t{c}^{2}+600\,{c}^{
2} \right) {T}^{2} \\
& & +4\, \left( 1+6\,\gamma_{{1}}{c}^{2} \right) t{c}^{2
}{X}^{4}+ \left( -24\, \left( 1+14\,\gamma_{{1}}{c}^{2} \right) {c}^{2
}t-600\,{c}^{2} \right) {X}^{2}  \\
& & +4800\, \left( 1+6\,\gamma_{{1}}{c}^{2}
 \right) {c}^{4}tX-12\, \left( 5+46\,\gamma_{{1}}{c}^{2} \right) {c}^{
2}t-600\,{c}^{2} ,\\
&H_{2trig}=&  {T}^{6}+3\,{X}^{2}{T}^{4}+ \left( 12\, \left( 86\,\gamma_{{1}}{c}^{2}+
9 \right) t{c}^{2}+1200\,{c}^{2} \right) {T}^{3} \\
& & + \left( 3\,{X}^{4}+
3600\,{c}^{2}X \right) {T}^{2} - \left( 3600\,{c}^{2} + 72\, \left( 22\,
\gamma_{{1}}{c}^{2}+1 \right) t{c}^{2} \right) {X}^{2}T \\
& & +{X}^{6}+3\,{X}^{4}+27\,{X}^{2}+14400\,{c}^{4} \left( 34\,\gamma_{{1}}{c}^{2}+3 \right) t \\
& & +144\, \left( 11+228\,\gamma_{{1}}{c}^{2}+1228\,{\gamma_{{1}
}}^{2}{c}^{4} \right) {t}^{2}{c}^{4}+720000\,{c}^{4} \\
& & -1200\,{X}^{3}{c}^{2}+3600\,{c}^{2}X+9 ,
\end{eqnarray*}
and X,T are defined in Eq. (8). Figure \ref{fig6} is plotted for the
$|q^{[2]}_{rwtrig}|^2$ to show its compression effect.  Because of
the explicit appearance of $t$ in $F_2, G_2, H_2, F_{2trig}$,
$G_{2trig}, H_{2trig}$, $q^{[2]}_{rw}$,  and $q^{[2]}_{rwtrig}$ one is
not be able to derive from the corresponding RWs of the NLSE by a mere scalar transformation of $x,t$.

Next, we construct the third-order RW of GNLSE by substituting
present $f_i(i=1,2,\cdots, 6)$ in $q^{[3]}$ \cite{hezhangwangpf2012}.
There exists  three patterns: a fundamental pattern
$q^{[3]}_{rw}$ with $s_1=s_2=0$, a triangular pattern
$q^{[3]}_{rwtrig}$ with $a=s_2=0$, and a circular pattern
$q^{[3]}_{rwcirc}$ with $a=s_1=0$. The explicit form $q^{[3]}_{rw}$
of the fundamental pattern with the third-order RW is given in the
Appendix, and, for brevity,  the very lengthy forms of the other two
cases are deleted.  Note that $q^{[3]}_{rw}$ includes the $t$-dependence explicitly. This fact shows that the second-order RW
solution cannot be obtained from the corresponding solution of the
NLS equation by a scalar transformation of $x,t$. Furthermore, to
show the compression effect on the RWs, Figs.
(\ref{fig7}-\ref{fig9}) are plotted for different parametric
choices.

\section{Conclusions}
In this paper, we considered the integrable version
of the generalized nonlinear Schr\"odinger equation with
several higher-order nonlinear terms, which describes ultra short pulse propagation through nonlinear silica fiber and soliton-type nonlinear excitations in classical Heisenberg spin chain. Using Daurboux transformation
and periodic seed solutions, we have constructed the first-order
breather solution and also discussed the behavior of these solutions with an
infinitely large period. Finally, we have also constructed the first-order,
second-order and third-order rogue wave solutions by the Taylor expansion.
All of these solutions have parameter $\gamma_1$ denoting the contribution
of higher-order nonlinear terms. The compressed effects of these solutions
are discussed through numerical plots by increasing the value of
$\gamma_1$.
This new phenomenon of the rogue wave is useful for us to observe or analyze its evolution in some complicated physical system. Another advantage of our results of this paper is that, as GNLSE is equivalent to spin chain, the rogue wave nature of spin systems can also be explained through suitable geometrical and gauge equivalence methods. In addition, as higher-order linear and nonlinear effects in optical fibers are playing key roles in explaining the generation and propagation of ultra short pulse through  silica wave guides,
we hope that our results with all these higher-order effects can be observed in
real experiments in the near future.

\section*{Acknowledgments}

This work is supported by the NSF of China
under Grants No. 10971109 and No. 11271210, K. C. Wong Magna Fund in Ningbo
University. J.H. is also supported by the Natural Science
Foundation of Ningbo under Grant No. 2011A610179. J.H. thanks
A. S. Fokas for his support during his visit at Cambridge.  K.P. thanks the DST, DAE-BRN, and
CSIR, Government of India, for the financial support through major
projects. L.W. is also supported by the Natural Science
Foundation of China, under Grants No. 11074136 and No. 11101230, and the
Natural Science Foundation of Zhejiang province under Grant No.
2011R09025-06.

\section*{Appendix: THE THIRD-ORDER ROGUE WAVE}
{\footnotesize
\begin{appendix}
$$q^{[3]}_{rw}=\left( \frac{24(F_3+iG_3)}{H_3}   + 1 \right) c{{\rm e}^{2\,i{c}^{2} \left( 3\,\gamma_{{1}}{c}^{2}+1 \right) t}}$$
\begin{eqnarray*}
&F_3=&11\,{T}^{10}+45\,{X}^{2}{T}^{8}+180\,
\left( 130\,\gamma_{{1}}{c}^{2}+
11 \right) t{c}^{2}{T}^{7}+ \left( 70\,{X}^{4}+420\,{X}^{2} \right) {T}^{6} \\
&&+ \left( 50\,{X}^{6}+480\, \left( 15300\,{\gamma_{{1}}}^{2}{c}^{4
}+2508\,\gamma_{{1}}{c}^{2}+73 \right) {t}^{2}{c}^{4} \right) {T}^{4}-
600\, \left( 38\,\gamma_{{1}}{c}^{2}+1 \right) t{c}^{2}{X}^{4}{T}^{3} \\
&&+ \left( 15\,{X}^{8}+7200\, \left( 2460\,{\gamma_{{1}}}^{2}{c}^{4}+308
\,\gamma_{{1}}{c}^{2}+15 \right) {t}^{2}{c}^{4}{X}^{2} \right) {T}^{2} \\
&&+ \left( -240\, \left( 38\,\gamma_{{1}}{c}^{2}+1 \right) t{c}^{2}{X}^{
6}-28800\, \left( 4056\,{c}^{6}{\gamma_{{1}}}^{3}+17+300\,{\gamma_{{1}
}}^{2}{c}^{4}+210\,\gamma_{{1}}{c}^{2} \right) {t}^{3}{c}^{6} \right) T \\
&&+{X}^{10}+15\,{X}^{8}+210\,{X}^{6}+ \left( -7200\, \left( 220\,{
\gamma_{{1}}}^{2}{c}^{4}+20\,\gamma_{{1}}{c}^{2}-1 \right) {t}^{2}{c}^
{4}-450 \right) {X}^{4} \\
&&+ \left( -43200\, \left( 628\,{\gamma_{{1}}}^{2
}{c}^{4}+124\,\gamma_{{1}}{c}^{2}+5 \right) {t}^{2}{c}^{4}-675
 \right) {X}^{2}+675 \\
&&+10800\, \left( 2452\,{\gamma_{{1}}}^{2}{c}^{4}+28
\,\gamma_{{1}}{c}^{2}-3 \right) {t}^{2}{c}^{4}
\end{eqnarray*}
\begin{eqnarray*}
&G_3=&{T}^{11}+5\,{X}^{2}{T}^{9}+20\, \left(
102\,\gamma_{{1}}{c}^{2}+5
 \right) t{c}^{2}{T}^{8}+ \left( 10\,{X}^{4}-60\,{X}^{2} \right) {T}^{7} \\
&&+ \left( 10\,{X}^{6}+480\, \left( 12\,{\gamma_{{1}}}^{2}{c}^{4}-268
\,\gamma_{{1}}{c}^{2}-29 \right) {t}^{2}{c}^{4} \right) {T}^{5}-120\,
 \left( 82\,\gamma_{{1}}{c}^{2}+7 \right) t{c}^{2}{X}^{4}{T}^{4} \\
&&+ \left( 5\,{X}^{8}+1440\, \left( 4524\,{\gamma_{{1}}}^{2}{c}^{4}+548\,
\gamma_{{1}}{c}^{2}+19 \right) {t}^{2}{c}^{4}{X}^{2} \right) {T}^{3}
+ \big( -80\, \left( 90\,\gamma_{{1}}{c}^{2}+7 \right) t{c}^{2}{X}^{6}\\
&&
-5760\, \left( 99432\,{c}^{6}{\gamma_{{1}}}^{3}+28676\,{\gamma_{{1}}}^
{2}{c}^{4}+3086\,\gamma_{{1}}{c}^{2}+107 \right) {t}^{3}{c}^{6}
 \big) {T}^{2} \\
&& + \left( -7200\, \left( 14\,\gamma_{{1}}{c}^{2}+1
 \right) ^{2}{t}^{2}{c}^{4}{X}^{4}+{X}^{10} \right) T-60\, \left( 14\,
\gamma_{{1}}{c}^{2}+1 \right) t{c}^{2}{X}^{8} \\
&&+120\, \left( 2\,\gamma_{
{1}}{c}^{2}-5 \right) t{c}^{2}{X}^{6}-1800\, \left( 10\,\gamma_{{1}}{c
}^{2}+3 \right) t{c}^{2}{X}^{4} \\
&&+ \left( 57600\, \left( -126\,\gamma_{{
1}}{c}^{2}+1176\,{c}^{6}{\gamma_{{1}}}^{3}-564\,{\gamma_{{1}}}^{2}{c}^
{4}-7 \right) {t}^{3}{c}^{6}+2700\, \left( 170\,\gamma_{{1}}{c}^{2}+7
 \right) t{c}^{2} \right) {X}^{2}\\
&&+18900\, \left( 14\,\gamma_{{1}}{c}^{
2}+1 \right) t{c}^{2}-14400\, \left( 11+1254\,\gamma_{{1}}{c}^{2}+
18084\,{\gamma_{{1}}}^{2}{c}^{4}+84168\,{c}^{6}{\gamma_{{1}}}^{3}
 \right) {t}^{3}{c}^{6}
\end{eqnarray*}
\begin{eqnarray*}
&H_3=&{T}^{12}+6\,{X}^{2}{T}^{10}+24\, \left(
206\,\gamma_{{1}}{c}^{2}+21
\right) t{c}^{2}{T}^{9}+ \left( 15\,{X}^{4}+270\,{X}^{2} \right) {T}^{8} \\
& &+ \left( 20\,{X}^{6}+720\, \left( 7596\,{\gamma_{{1}}}^{2}{c}^{4}+
1636\,\gamma_{{1}}{c}^{2}+83 \right) {t}^{2}{c}^{4} \right) {T}^{6}- 240\, \left( 42\,\gamma_{{1}}{c}^{2}-1 \right) t{c}^{2}{X}^{4}{T}^{5}\\
& &+ \left( 15\,{X}^{8}+8640\, \left( 3012\,{\gamma_{{1}}}^{2}{c}^{4}+492
\,\gamma_{{1}}{c}^{2}+25 \right) {t}^{2}{c}^{4}{X}^{2} \right) {T}^{4} \\
& & + \left( -240\, \left( 82\,\gamma_{{1}}{c}^{2}+3 \right) t{c}^{2}{X}^{
6}-57600\, \left( 3048\,{c}^{6}{\gamma_{{1}}}^{3}-2604\,{\gamma_{{1}}}
^{2}{c}^{4}-450\,\gamma_{{1}}{c}^{2}-17 \right) {t}^{3}{c}^{6} \right) {T}^{3}  \\
& &+ \left( 6\,{X}^{10}-21600\, \left( 548\,{\gamma_{{1}}
}^{2}{c}^{4}+76\,\gamma_{{1}}{c}^{2}+1 \right) {t}^{2}{c}^{4}{X}^{4} \right) {T}^{2}  \\
& &+ \big( 172800\, \left( 46968\,{c}^{6}{\gamma_{{1}}}^
{3}+11196\,{\gamma_{{1}}}^{2}{c}^{4}+906\,\gamma_{{1}}{c}^{2}+29
 \right) {t}^{3}{c}^{6}{X}^{2}\\
& & -360\, \left( 22\,\gamma_{{1}}{c}^{2}+1
 \right) t{c}^{2}{X}^{8} \big) T  \\
& &+{X}^{12}+6\,{X}^{10}+135\,{X}^{8}+
 \left( 2880\, \left( 1324\,{\gamma_{{1}}}^{2}{c}^{4}+164\,\gamma_{{1}
}{c}^{2}+3 \right) {t}^{2}{c}^{4}+2340 \right) {X}^{6} \\
& & + \left( -43200
\, \left( 428\,{\gamma_{{1}}}^{2}{c}^{4}-12\,\gamma_{{1}}{c}^{2}-5
 \right) {t}^{2}{c}^{4}+3375 \right) {X}^{4} \\
& & + \left( -64800\, \left(
2500\,{\gamma_{{1}}}^{2}{c}^{4}+492\,\gamma_{{1}}{c}^{2}+9 \right) {t}
^{2}{c}^{4}+12150 \right) {X}^{2}+2025 \\
& & +64800\, \left( 7260\,{\gamma_{{
1}}}^{2}{c}^{4}+772\,\gamma_{{1}}{c}^{2}+23 \right) {t}^{2}{c}^{4}  \\
& & + 172800\, \left( 213+8056\,\gamma_{{1}}{c}^{2}+720096\,{c}^{6}{\gamma_{
{1}}}^{3}+115128\,{\gamma_{{1}}}^{2}{c}^{4}+1836624\,{\gamma_{{1}}}^{4
}{c}^{8} \right) {t}^{4}{c}^{8}
\end{eqnarray*}
\end{appendix}
}



\setcounter{figure}{0}
\renewcommand{\thefigure}{\arabic{figure}}
\begin{figure}[!h]
\centering
\subfigure[$a=\xi=0,c=\frac{2}{5},\eta=\frac{1}{2},\gamma_1=0$]{\includegraphics[scale=0.35]{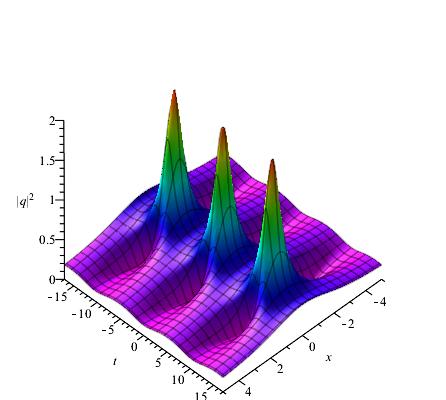}}
\subfigure[$a=\xi=0,c=\frac{2}{5},\eta=\frac{1}{2},\gamma_1=\frac{1}{2}$]{\includegraphics[scale=0.35]{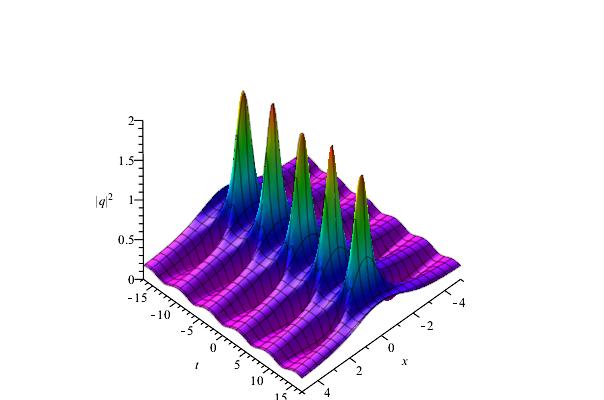}}
\subfigure[$a=\xi=0,c=\frac{2}{5},\eta=\frac{1}{2},\gamma_1=1$]{\includegraphics[scale=0.35]{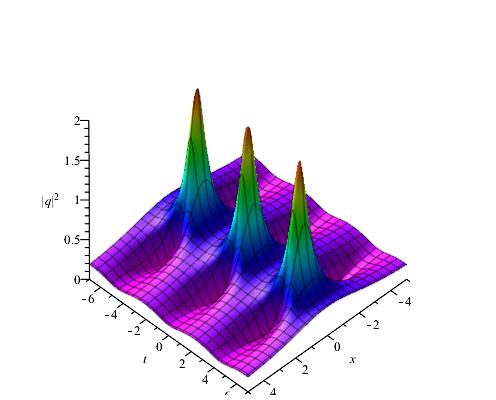}}
\subfigure[$a=\xi=0,c=\frac{2}{5},\eta=\frac{1}{2},\gamma_1=\frac{3}{2}$]{\includegraphics[scale=0.35]{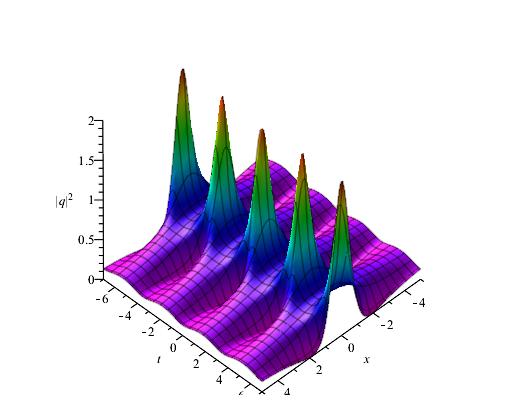}}\\
\subfigure[$a=\xi=0,c=\frac{2}{5},\eta=\frac{1}{2},\gamma_1=2$]{\includegraphics[scale=0.3]{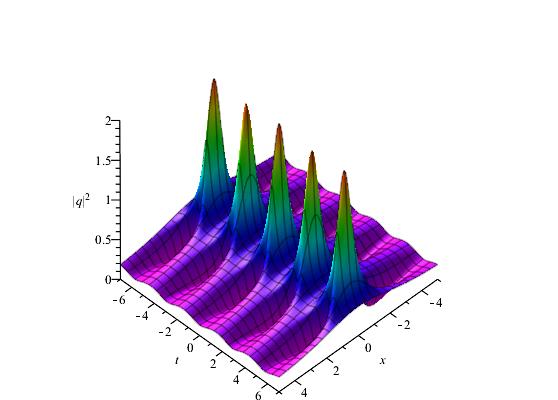}}
\subfigure[$a=\xi=0,c=\frac{2}{5},\eta=\frac{1}{2},\gamma_1=3$]{\includegraphics[scale=0.3]{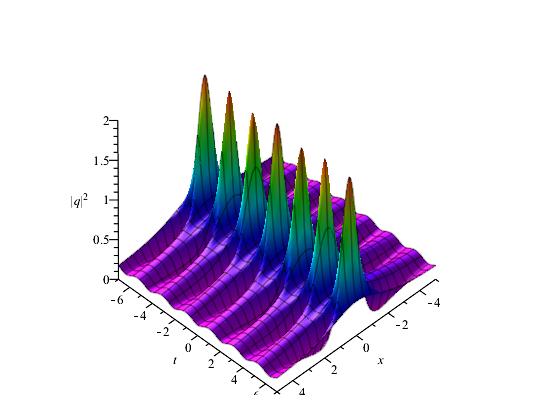}}
\caption{(Color online) The dynamical evolution of the first-order breather $|q^{[1]}|^2$ on
the ($x,t$) plane. When the value of $\gamma_1$ increases, the number of peaks on same interval of time
also increases.}
\label{fig2}
\end{figure}

\begin{figure}[!h]
\centering
\subfigure[$c=\frac{1}{2},\gamma_1=0$]{\includegraphics[scale=0.3]{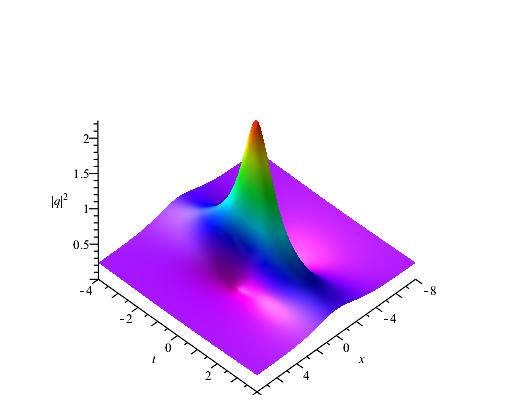}}
\subfigure[$c=\frac{1}{2},\gamma_1=\frac{1}{2}$]{\includegraphics[scale=0.3]{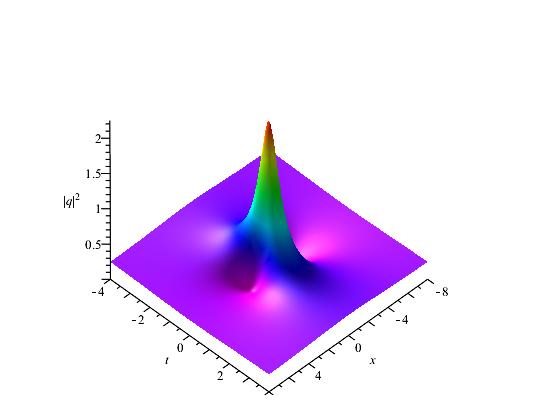}}\\
\subfigure[$c=\frac{1}{2},\gamma_1=1$]{\includegraphics[scale=0.3]{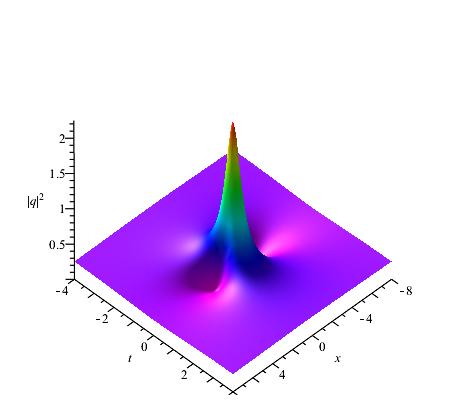}}
\subfigure[$c=\frac{1}{2},\gamma_1=2$]{\includegraphics[scale=0.3]{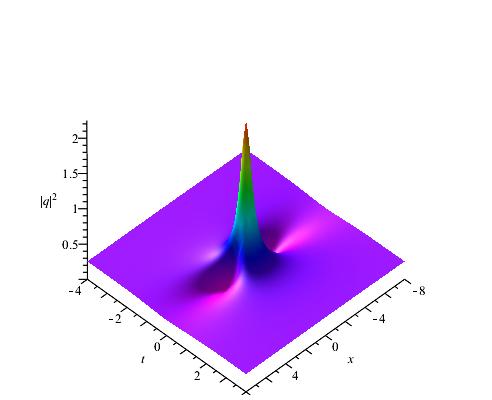}}
\caption{(Color online)
The dynamical evolution of the first-order rogue wave
$|q^{[1]}_{limit}|^2$ on the ($x,t$) plane.
For larger values of $\gamma_1$, it is clear that the compression effects in $t$ direction are quite high. }
\label{fig4}
\end{figure}

\begin{figure}[!h]
\centering
\subfigure[$a=s_1=0,c=\frac{1}{\sqrt{2}},\gamma_1=0$]{\includegraphics[scale=0.4]{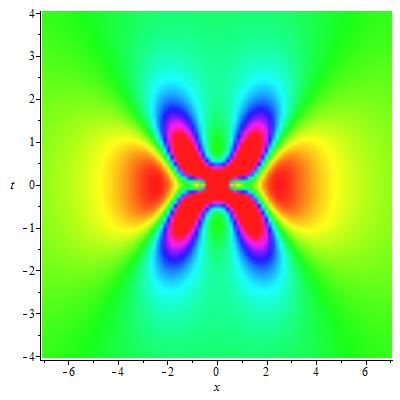}}
\subfigure[$a=s_1=0,c=\frac{1}{\sqrt{2}},\gamma_1=0$]{\includegraphics[scale=0.4]{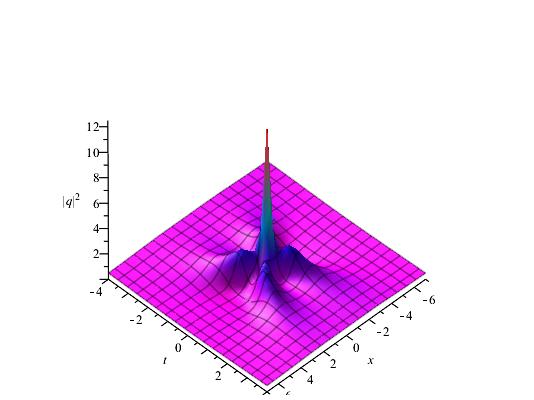}}\\
\subfigure[$a=s_1=0,c=\frac{1}{\sqrt{2}},\gamma_1=1$]{\includegraphics[scale=0.4]{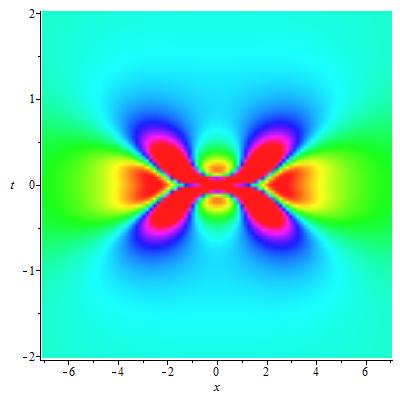}}
\subfigure[$a=s_1=0,c=\frac{1}{\sqrt{2}},\gamma_1=1$]{\includegraphics[scale=0.4]{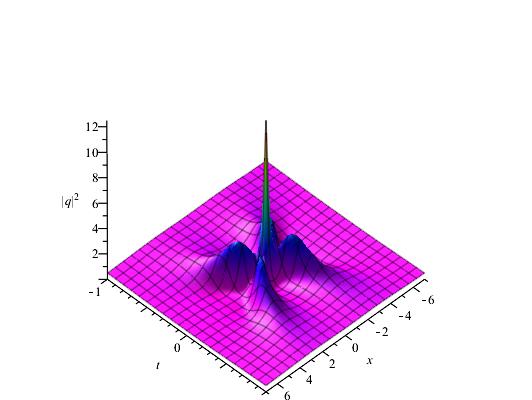}}
\caption{(Color online)
The dynamical evolution of the second-order rogue wave
$|q^{[2]}_{rw}|^2$ on the ($x,t$) plane. Comparing  (a) and (b)
with (c) and (d) indicates effective high
compression in the $t$ direction.}
\label{fig5}
\end{figure}

\begin{figure}[!h]
\centering
\subfigure[$a=0,c=\frac{1}{\sqrt{2}}, s_1=50-50i,\gamma_1=0$]{\includegraphics[scale=0.3]{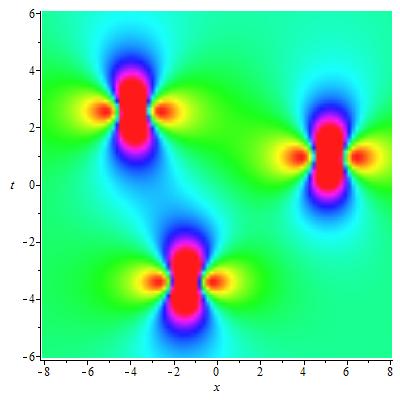}}
\subfigure[$a=0,c=\frac{1}{\sqrt{2}},s_1=50-50i,\gamma_1=0$]{\includegraphics[scale=0.4]{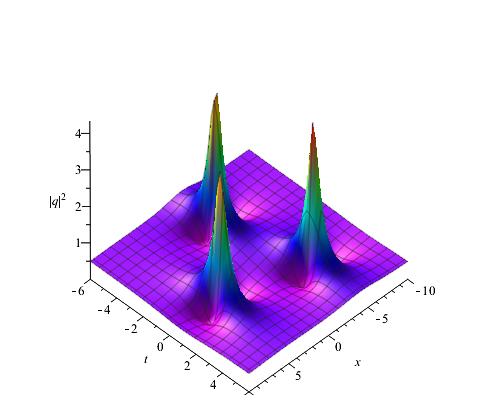}}\\
\subfigure[$a=0,c=\frac{1}{\sqrt{2}},s_1=50-50i, \gamma_1=\frac{1}{4}$]{\includegraphics[scale=0.4]{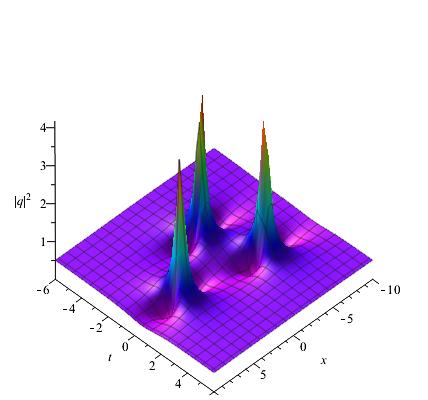}}
\subfigure[$a=0,c=\frac{1}{\sqrt{2}},s_1=50-50i, \gamma_1=\frac{3}{4}$]{\includegraphics[scale=0.4]{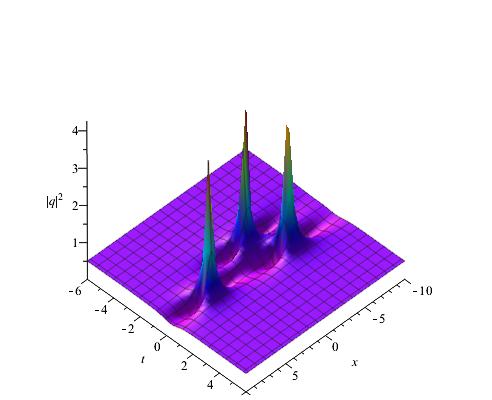}}
\caption{(Color online) The dynamical evolution of the second-order rogue wave
$|q^{[2]}_{rwtrig}|^2$ on the ($x,t$) plane. It is shown from (b),(c), and (d)
that rogue wave compression increases as the value of $\gamma_1$ increases.
}
\label{fig6}
\end{figure}

\begin{figure}[!h]
\centering
\subfigure[$a=s_1=0,c=\frac{1}{\sqrt{2}},\gamma_1=0$]{\includegraphics[scale=0.35]{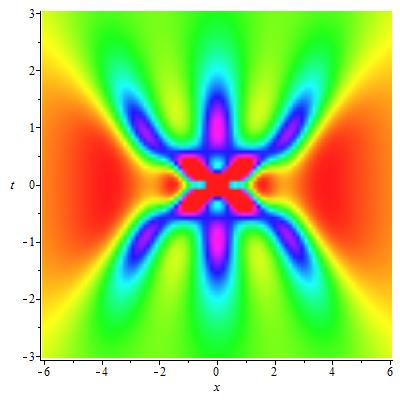}}
\subfigure[$a=s_1=0,c=\frac{1}{\sqrt{2}},\gamma_1=0$]{\includegraphics[scale=0.4]{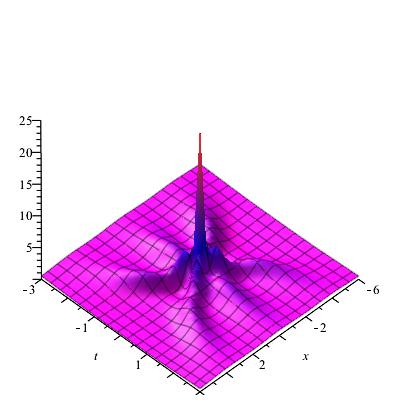}}\\
\subfigure[$a=s_1=0,c=\frac{1}{\sqrt{2}},\gamma_1=\frac{1}{4}$]{\includegraphics[scale=0.35]{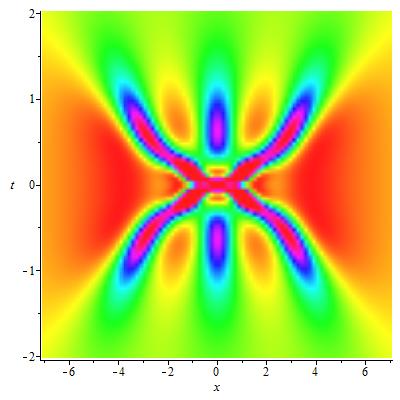}}
\mbox{\hspace{1cm}}\subfigure[$a=s_1=0,c=\frac{1}{\sqrt{2}},\gamma_1=\frac{1}{4}$]{\includegraphics[scale=0.4]{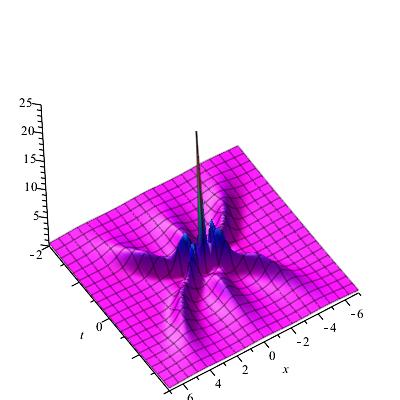}}
\caption{(Color online)
The dynamical evolution of the third-order rogue wave
$|q^{[3]}_{rw}|^2$ on the ($x,t$) plane.
By comparison with (a) and (b), (c) and (d) are highly compressed.}
\label{fig7}
\end{figure}
\begin{figure}[!h]
\centering
\subfigure[$c=\frac{1}{\sqrt{2}},s_1=-50i,\gamma_1=0$]{\includegraphics[scale=0.4]{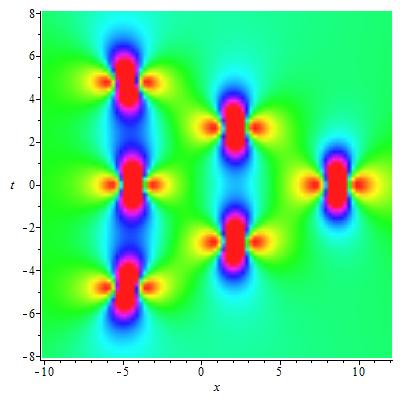}}
\subfigure[$c=\frac{1}{\sqrt{2}},s_1=-50i,\gamma_1=0$]{\includegraphics[scale=0.4]{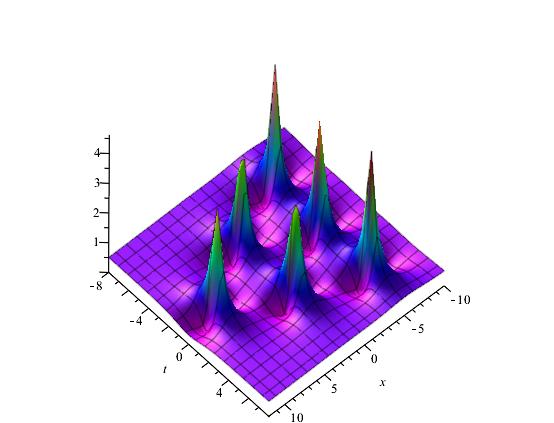}}\\
\subfigure[$c=\frac{1}{\sqrt{2}},s_1=-50i,\gamma_1=\frac{1}{4}$]{\includegraphics[scale=0.4]{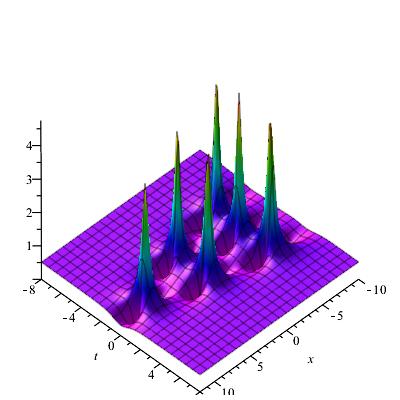}}
\subfigure[$c=\frac{1}{\sqrt{2}},s_1=-50i,\gamma_1=\frac{3}{4}$]{\includegraphics[scale=0.4]{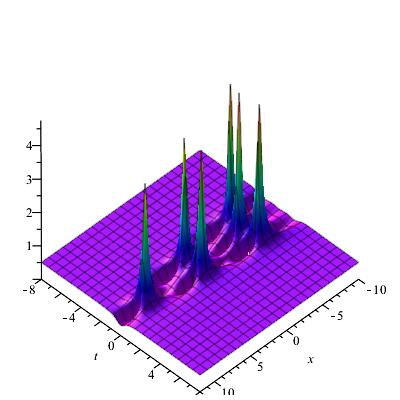}}
\caption{(Color online)The dynamical evolution of the third-order rogue wave
$|q^{[3]}_{rwtrig}|^2$ on the ($x,t$) plane. It is shown from (b), (c), and (d)
that rogue wave is compressed more while increasing the value of $\gamma_1$.}
\label{fig8}
\end{figure}
\begin{figure}[!h]
\centering
\subfigure[$a=0,c=\frac{1}{\sqrt{2}},s_2=5000i,\gamma_1=0$]{\includegraphics[scale=0.4]{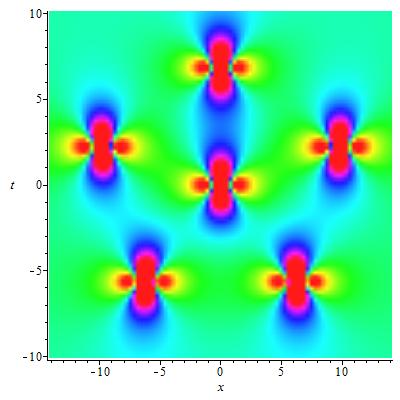}}
\subfigure[$a=0,c=\frac{1}{\sqrt{2}},s_2=5000i,\gamma_1=0$]{\includegraphics[scale=0.4]{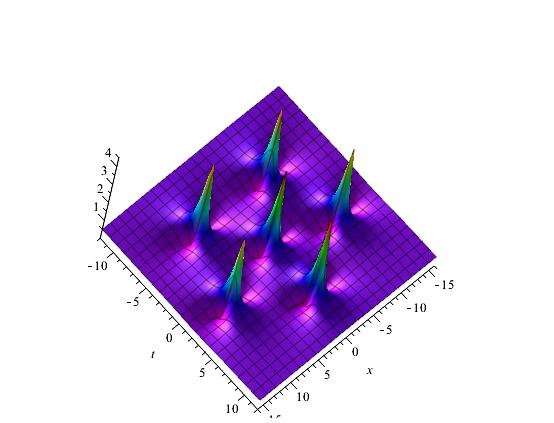}}\\
\subfigure[$a=0,c=\frac{1}{\sqrt{2}},s_2=5000i,\gamma_1=\frac{1}{4}$]{\includegraphics[scale=0.4]{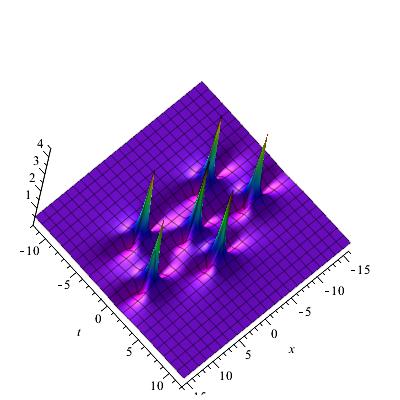}}
\mbox{\hspace{1cm}}\subfigure[$a=0,c=\frac{1}{\sqrt{2}},s_2=5000i,\gamma_1=1$]{\includegraphics[scale=0.4]{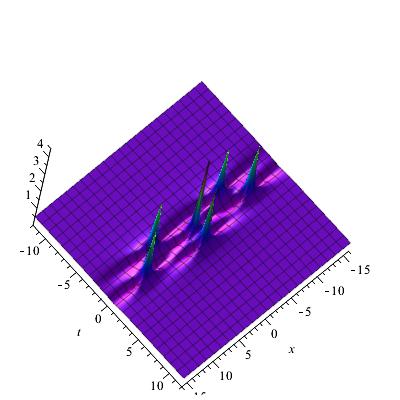}}
\caption{(Color online)
The dynamical evolution of the third-order rogue wave
$|q^{[3]}_{rwcirc}|^2$ on the ($x,t$) plane. It is shown from (b), (c), and (d)
that rogue wave is changing its shape and also compression increases  by increasing the value of $\gamma_1$.}
\label{fig9}
\end{figure}

\end{document}